\documentstyle[12pt,fullpage]{article}

\begin{document}
\begin{center}
{\large\bf New Vortex with Zero Fluxoid Quantum in a Superconductor :}\\
{\large\bf Anomalous Hall Effect in High T$_c$ and Low T$_c$ Superconductors}
\vskip1cm

 Sang Boo Nam\\
{\sl University Research Center, 7735 Peters Pike, Dayton, OH 45414 USA}\\
and\\
{\sl Department of Physics, Pohang University of Science and Technology}\\
{\sl Pohang, Kyungbook 790-784, KOREA$^*$}
\vskip2cm

{\bf abstract}
\end{center}
\noindent
It is shown that  there are three kinds  of vortices in a  superconductor; a vortex
 with  the fluxoid
quantum 1 (VF), a new vortex with 0 (VZF), and a vortex with $-1$  (VAF), respectively.
The Hall
field via three vortices is studied and found to change its sign, depending of the
relative strength of 
contributions via VF and VAF,  similar to that of  a system with two  kinds of carriers.
VZF  is 
suggested to act  as a domain  wall between  VF and VAF  regions, accounting  for the
flux  and
antiflux regions observed in Nb films.
\vspace*{1cm}

keywords: superconductivity, vortex, zero fluxoid quantum, Hall effect\\
$^*$e-mail : sbnam@galaxy.postech.ac.kr
\newpage
     
Vortices play  important roles   for our understanding  of  the properties  of  superconductors.  
However, there  has been   no discussion, hitherto,   of all possible   fluxoid quantum numbers   of 
vortices.  Here, I discuss the fluxoid quantum states of  vortices and their applications to the Hall 
anomaly.

Recently, it is shown [1],  within the BCS [2]  pairing theory, that to  have a finite transition 
temperature T$_c$ in a superconductor, a  finite pairing interaction range T$_d$ is  required, and that not 
all carriers participate in  pairings, yielding multiconnected  supereconductors (MS).  The region  of 
unpaired carriers in a MS [3] may be considerd as a vortex with anti-flux (VAF) [flux(VF)], when 
the local  magnetic  flux around   a vortex,  is less   [greater] than  the equivalent   flux via  the 
supercurrent.  In other words, VF and VAF are vortices with the London fluxoid quanta 1 and $-1$, 
respectively.  Naturally, one may consider a vortex with zero fluxoid quantum  (VZF), when the local 
magnetic flux is  balanced with  the equivalent  flux via  the supercurrent [in other  words, the 
current(j) goes with the vector  potential(A) as $j \propto -A$  near the vortex core (the  London gauge)]. 
Indeed, I show later there are three kinds of vortices, analogous to  three magnetic quantum states 
of the angular momentum 1.  On the other hand, thermally activated  vortex-antivortex [4] bindings 
below the  Kosterlitz-Thouless temperature  T$_{KT}$ [5]  and dissociations   above T$_{KT}$ are  not to  be 
discussed here.

    The motion of  a vortex  is known  to yield  the Hall  field in  superconductors.  The  force 
responsible for the vortex motion is  still controversial, the Lorentz force on  the vortex core [6] or 
the Magnus force [7].  The latter is shown to be present in a superconductor  [8].  The data [9] of 
the resistive transition initially incompatible with the  Lorentz force can be understood in the  frame 
work of  the  flux  flow [10].   A scaling  behavior of  the Hall   resistivity
$\rho_{yx} \propto \rho^2_{xx}$,  is 
understood in the vortex state with  disordered dominated dynamics [11], compatible  with the flux 
flow picture [6]. But, both pictures of [6, 7] are not able to account for the sign changes of the Hall 
field during sweeping temperature T or  magnetic field H; once [12],  twice [13, 14] and three  times 
[15].  To account for the sign change of the Hall field, there are proposals  of the force via pinning 
centers [16] and the opposing drift of the thermally excited quasiparticles near T$_c$ [17].  The former 
is incompatible with data [13]  and the latter with  data [12-15] at low  T.  A charge at  a vortex 
core [18] opposite  to carrier  charge, yielding  a force  opposite to  the Lorentz  force, is  unlikely 
possible, due to the supercurrent.  On the other hand, the time dependent Ginzberg Landau equation 
approach [19]  requires  the imaginary   part of  the relaxation   time, reflecting  the particle  hole 
asymmetry.  A proposal of breaking the particle hole symmetry [20],  in the range of the coherence 
length being  of the  order of   the mean free  path, is  hardly  realizable, since  the particle  hole 
symmetry depends on neither  the coherene length nor  the mean free path.   So far, one quantum 
vortex state, VF,  has been  examined via  various proposals.   It is  desirable to  explore a new 
approach, taking into account all possible quantum states of vortices.  
     
In this paper,  I show  that there are  three kinds  of vortices; VF,  VZF, and  VAF, via the 
Ginzburg Landau (GL) equation, and that the sign of the Hall field depends on the relative strength 
of contributions  via VF   and VAF, assuming   the vortex motion   without addressing the   force 
responsible for it.  Here natural units of $\hbar = c = k_{B} = 1$ and the flux unit
$\Phi_0=hc/2e=1$ are used. 
     
Macroscopically, the current in a superconductor may be given by
\begin{eqnarray}
j(r) = j_o(r)[\nabla \theta(r)/2\pi-A(r)/(hc/2e)],
\end{eqnarray}
where $\theta (r)$ is the phase of order  parameter, $A(r)$ is the vector potential,  and $j_o(r)$ is the square of 
the order  parameter amplitude   times charge/mass $(2e/2m)$.    The London fluxoid   quatnum (n) 
condition may be obtained as 
\begin{eqnarray}               
n=\oint\nabla \theta(r)/2\pi \cdot dl = \Phi_A+\Phi_J,
\end{eqnarray}
where $\Phi_A$ is the magnetic  flux enclosed by the  integral contour, $\Phi_J$ is  the effective flux via  the 
line integral of  $j(r)/j_o(r)$, and the  integer n is  a reflection of  the single valuedness  of the order 
parameter.  Let us write the order  parameter in the cylindrical coordinates with  the origin where 
the order parameter vanishes, $\Psi(r, \varphi) = |\Psi(\infty)| f(r) \exp$ (i n $\varphi$), omitting  the z-dependence.  The 
GL and Maxwell's equations are given as [21]
\begin{eqnarray}
&&\xi^2[(d/rdr)(rd/dr) - q(r)^2]f(r)+f(r)=f(r)^3,\\
&&(d/dr)(d/rdr)[rq(r)]=-8\pi^2 j(r)=q(r) f(r)^2/\lambda^2,
\end{eqnarray} 
where $q(r) = n/r - 2\pi A(r)$ is the superfluid  velocity, $A(r)$ and $j(r)$  are the vector potential  and 
current in the azimuthal direction, $\lambda$ is the effective pentration depth length $\lambda^2=2m/4\pi | 2e\Psi(\infty)|$ 
and $\xi$ is  the coherence length.   The magnetic  field in the  z-direction is 
$h(r) = (d/rdr)[rA(r)]$.
From Eqs. (3) and (4), we can see that once $A(r)$ for VZF (n = 0) is known, $q(r)$ is obtained,  and 
vice versa.  Thus, three kinds of vortices  with n = 1, 0 and  $-1$ are possible in a MS.   However, 
VZF is not possible in a singly connected superconductor, since a vortex is formed after a magnetic 
flux entered, that is, n =  1.  To find a solution of  $A(r)$ for VZF explicitly, let me  consider $f(r) = 
(r/r_c)^p$ for $r < $ the effective  core size $r_c$ which will be  determined later by the variational method.   
Solving Eq.  (4) for n = 0, we obtain $A(r)$ via the Bessel function with a pure imaginary argument 
as [22]
\begin{eqnarray}
A(r) \propto K_u(Z), {\rm ~with}~ u = 1/(1+p) {\rm ~and~} Z=u(r_c/\lambda)(r/r_c)^{1+p}.
\end{eqnarray}
Eq.  (5) is also  a solution for $q(r)$  for any n and  its elaboration will be  reported elsewhere [22].   
For p = 0, Eq.  (5) yields a solution $K_1(r/\lambda)$ of  the London equation, and is not acceptable near a 
vortex core where the order parameter vanishes.  For $p = 1, Z = r^2/2\lambda r_c$,  we get $A(r)$, and $j(r)$ as
\begin{eqnarray}
A(r) &=& -[\phi(0)/2\pi r]\exp(-Z), {\rm ~with}~\phi(0)=2\pi h(0)\lambda r_c,\\
h(r) &=& h(0) \exp (-Z),\\
j(r) &=& -[f(r)/\lambda]^2 A(r)/4\pi = [h(0)/4\pi \lambda r_c] r\exp (-Z).
\end{eqnarray}
The $h(r)$ has a Gaussian distribution and $j(r)$ has a maximum at $r = (\lambda r_c)^{1/2}$.  Thus, a vortex  may 
be considered as a superconducting  ringlet.  Inserting Eq.  (6) into  Eq.  (3) for $f(r)$, we  find the 
condition
\begin{eqnarray}
\phi (0) = 2\pi h(0) \lambda r_c = {\rm~ a~flux~unit}
\end{eqnarray}
to have a solution of $f(r)$  consistently.  Eq. (9) provides a key  for having three kinds of vortices, 
together with the idea of superconducting ringlets.  Furthermore, we rewrite $A(r)$ of Eq.  (6) near $r = 0$
 as
\begin{eqnarray}
A(r) = h(0)r/2 - \phi(0)/2\pi r .
\end{eqnarray}
The first term is like a vector potential of a magnetic field $h(0)$ and the second indicates the fluxoid 
quantum number n = $-1$, in opposing direction to $h(0)$.  This  confirms the notion of the circulating 
supercurrent guarding the normal region from a small  magnetic field [3].  In other words, a vortex 
acts like a superconducting ringlet.   For VF and VAF,  $A(r)$'s may be given  as $A(r) \propto (\pm) [1 - \exp(-Z)]/2\pi r$, 
where + for VF and $-$ for VAF.  $j(r)$'s may be written like Eq.  (8) with the proper 
singns such that Eq.  (2) is satisdied.
  
To determine $r_c$, let me consider the variational function
\begin{eqnarray}
f(r) = {\rm tanh}(r/r_c)
\end{eqnarray}
and $A(r)$ of Eq. (6).  The GL free energy density $F$ associated  with a vortex, measured relative to 
the free energy of the Meissner state, in the dimensionless form, may be given as [21]
\begin{eqnarray}
F=[1-f^2(r)]^2/2+[df(r)/\kappa dr]^2+[A(r)f(r)]^2+h^2(r),
\end{eqnarray}
where the GL parmeter $\kappa  =\lambda / \xi$, the dimensionless quantities  are $F$ in the units  of $H_c^2/4\pi$ with 
the thermodynamical critical field $H_c,~h(r)$ in the units of $\sqrt{2} H_c,~ A(r)$ in the units of $\sqrt{2} H_c \lambda$, and 
$r$ in the units of $\lambda$.  By integrating $F$, we may obtain the energy per unit length of vortex line, or 
equivalently the low critical field $H_{c1}$ as
\begin{eqnarray}
H_{c1}/ \sqrt{2} H_c = (\kappa /2) \int^\infty_0 {\rm ~rdr~} F.
\end{eqnarray}
After carrying out all integrals in Eq. (13) and some algebra of minimizing  it with respect to $r_c$, 
we may obtain
\begin{eqnarray}
H_{c1}/\sqrt{2} H_c &=& [D(3z^2+2)+E_1(z/\kappa)]/4\kappa,\\
E_1(z/\kappa) &=& \int^\infty_1 dx \exp (-xz/\kappa)/x,\nonumber
\end{eqnarray}
where $D = 0.29543$, and $z = r_c/\xi$ is  determined by $z^3 = [ 2 - \exp(-z/\kappa)]\kappa/2D$.   For $k = 0.707$, we 
obtain $r_c = 1.3\xi$, and $H_{c1}/H_c = 1.042$ about 4\% higher  than the exact GL result of 1, and $h(0)/H_c 
= 1.088$ about 9\% higher than the exact GL  result of 1.  The present model of Eqs.  (11)  and (6), 
appears to be reasonable.  The  variational function [23] $f(r)  = r/[r^2 + r_c^2]^{1/2}$ has been studied and 
yields the results in  good agreements with the  numerical solutions [24] of  the GL equations.   In 
fact, the numerical field solution [24] has a form of Eq.  (7) which leads me consider vortices (flux 
lines) as noninteracting particles for later discussion of the Hall field.  

     Let us pause a moment and consider the energy of the n flux  state in a superconducting ring, 
$E(n)\propto(n-\Phi_A)^2$. At $\Phi_A = n + 1/2$, the energies  of $n$ and $n + 1$ states are same, $E(n)  = E(n +1)$, 
but the currents in n and  n + 1 states flow  in opposite directions, to staisfy the  fluxoid quantum 
condition.  The fluxoid quantum  has been demonstrated  experimentally some time  ago [25].  We 
argue the same reasoning be applicable to a vortex, via $\phi(0)$ being $\Phi_A$,  by considering a vortex as 
a superconducting ringlet as metioned before.  Then, making use of a flux unit, $1 = 2\pi\sqrt{2} H_c \lambda\xi$, 
and shifting a center for VZF by 1 flux unit via Eq.  (9), we may have, with $z = r_c/\xi$, 
\begin{eqnarray}
&&{\rm VF~ for~} 3/2 < [\phi(0) {\rm~or~} h(0)z/\sqrt{2} H_c] < 5/2,\nonumber\\ 
&&{\rm VZF~ for~} 1/2 < [\phi(0) {\rm ~or~} h(0)z/\sqrt{2} H_c] < 3/2,\nonumber\\
&&{\rm VAF~ for~} 0 < [\phi(0) {\rm~or~} h(0)z/\sqrt{2} H_c] < 1/2,
\end{eqnarray}
respectively. $\phi(0)$ can not be $> 5/2$, because a vortex can have the fluxoid quantum $< 2$, unlike a 
superconducting ring, via the  magnetic energy consideration.   Physically, the fluxoid  quanta 1, 0, 
and $-1$ for VF,  VZF and VAF,  are equivalent to  the magnetic quantum  numbers of the angular 
momentum 1.  Note  that all  variables $h(0)$, $\lambda, ~\xi$ should  be understood  as the  local variables 
depending on the local scattering centers and mean order parameter.  
     Therefore, VF, VZF  and VAF  may be formed  at different  regions, depending on  the local 
conditions.  Let us consider VF and  VAF far apart each other.   The circulating currents between 
them, flow in the  same direction, so that  they are not collapsed.   When VF and  VAF are close 
together, they may or may not be collapsed, depending on  the local conditions.  We may think of 
a domain wall made  of VZF which  divides the regions of  VF and VAF,  similar to the  magnetic 
domain wall.  The present scenario confirms the prediction [3] of VAF as observed in niobium films [26] and YBCO [27]. The key idea is that regions of unpaired carriers [1] act 
as vortices [3].   In the symmetry  point of view,  the global periodicity  is broken to  have VZF 
acting as a domain wall between VF and VAF  regions where the local periodicities are maintained, 
compatible with data  of VF and  VAF in  Nb films [26].   In other  words, the local  free energy 
minimum may play a role for the vortex states, rather than the global free energy minimum.  Note that 
when a transport current flows  in a system, then conditions  for VAF and VZF may  be enhanced 
via an additional  circulating current induced  by the  transport current and  magnetic field. Here  no attempt is made to treat the magnetic field of transport current self-consistently.
  
     Now we examine  the Hall  field via  VF, VZF,  and VAF  by considering  them like  stable 
particles with the magnetic quantum numbers 1, 0, and $-1$, respectively.  Let the transport current J 
be in the x-direction and  the magnetic filed H  in the z-direction of  unit vector ${\bf k}$.  We  consider 
here the motions of vortices only.  The momentum balance may be written as
\begin{eqnarray}
J = \displaystyle{\sum_i}[a_i V_{Li}+ b_i V_{Li}\times {\bf k}]
\end{eqnarray} 
with the i-vortex velocity $V_{Li} = (x_i V_x, y_i  V_y, 0)$, where $i = 1$ for VF, 2  for VAF and 3 for VZF.  
The parameters $a = (a_1, a_2, a_3)$ and $b = (b_1, b_2, b_3)$ are to be determined later.  Physically, $x_i$ and $y_i$ 
may be considered as the mean values of velocities of same all vortices  in the units of $V_x$ and $V_y$.  
The electric field induced by a  moving vortex with a flux $\Phi_i$, is known to be  $-V_{Li} \times  {\bf k}\Phi_i$ [28].   
The total electric field induced by all vortices, may be given by 
\begin{eqnarray}
E = -\displaystyle{\sum_i}V_{Li} \times {\bf k}\Phi_i= {\bf \rho} J
\end{eqnarray}
where ${\bf \rho}$ is the flux flow resistivity tensor and $\Phi=(\Phi_1, \Phi_2, \Phi_3)$ with the i-flux in the z-direction 
$\Phi_i$ of the fluxoid quantum $n$ times the i-vortex number $n_i$.  From Eqs.  (16) and (17), we may get 
the resistivity tensor and the Hall angle $\theta_H$ as 
\begin{eqnarray}
\rho_{xx} &=& (y\Phi)(bx)/[(ax)(ay)+(bx)(by)],\\
\tan \theta_H &=& \rho_{yx}/ \rho_{xx}=(x\Phi)(ay)/(y\Phi)(bx),
\end{eqnarray}
where $(AB)=\displaystyle{\sum_i} A_i B_i$ .  By choosing $V_x = V_T = V_y$, we get $\rho_{xx} = (y\Phi)/eN$ with $eN = J/V_T$ and
$\tan \theta_H = (x\Phi)/(y\Phi)$.  By considering noninteracting vortices, $a_i$ and $b_i$ may be obtained by the basic 
equations of [6, 7],
\begin{eqnarray}
V_{Li} \times {\bf k} &=& h_iV_T \times {\bf k} + V_T / W_i,\\
V_T e N_i &=& a_i V_{Li} + b_i V_{Li} \times {\bf k},\nonumber\\\
a_i / W_i h_i &=& b_i = e N_i W_i / [1+(W_i h_i)^2],
\end{eqnarray}
where $V_T$ is  the velocity  of the transport  carrier, $N_i$  is the carrier  number associated  with the 
i-vortex, and $\displaystyle{\sum_i} N_i = N =$ the carrier density.   $h_i = ~< B > = B(T, H)/H_{c2} = [H -  M(T, H)]/H_{c2}$ with 
the magnetization $-M(T, H)$, for the Lorentz force on the  vortex core [6] and $h_i =1$ for the Magnus 
force [7].  $W_i =  eH_{c2} \tau_i/m = W_c \tau_i$ with the carrier scattering time $\tau_i$ in the i-vortex core.  VZF 
moves with $V_T$ and yields  no electric field in  the transport direction, resulting in no dissipation of energy and no flux flow resistivity, equivalently,  $W_3 =\infty$.  For 
VAF, $H_{c2}$ should be $-H_{c2}$, because the circulating supercurrent is equivalent to  $-H_{c2}$.  Let us consider a case of 
$x = (h_1, h_2, 1)$ and $y  = (1/\tau_1, -1/\tau_2, 0)/W_c$ with $V_x  = V_T= V_y$, via Eq.   (20).  From Eqs.  
(19) and (21), we may obtain
\begin{eqnarray*}
~~~~~~~~~~~~~~~~~~~~~~~~~~~~~~~~\tan \theta_H = W_c[n_1 h_1 - n_2 h_2] / [ n_1/ \tau_1 + n_2 /\tau_2].\hspace*{3cm}({\rm 19A})
\end{eqnarray*}
Now, for $n_2/n_1$, I recapitulate the argument used to predict the sign changes of the Hall field three 
times [3] before the experiment [15] is known.  For VF, $n_1 = AB(T,  H)/\Phi_0$, with a sample area A.  
Since VAF  is  formed mainly  by  the circulating  supercurrent,  $n_2$ may  be  proportional to  the 
superelectron density $N_s(T, H)$ times the normal fraction,  $1~ - N_s(T, H)/N = 1 - <N_s>$.   We may 
get, $n_2 = PSA/\pi \xi^2$, with  $S = ~<N_s>[1~ - <N_s>]$.   P is a parameter function  of T, H and J,  and 
zero for J = 0 and H = 0, since the  circulating supercurrent is induced by J and H.  In fact,  $n_2$ is 
PS times total carriers in  a sample/carriers inside a  vortex volume.  Then, $n_2/n_1  = 2PS/<B>$ with 
$<B>~ = B(T, H)/H_{c2}$, we may get 
\begin{eqnarray*}
~~~~~~~~~~~~~\tan \theta_H = W_c [<B> h_1 - 2PS h_2 ]/ [<B> /\tau_1 + 2PS/\tau_2].\hspace*{3cm}({\rm 19B})
\end{eqnarray*}
2S has a  finite value at  T = 0  and  H =  0, scince unpaired  carriers are found  to be 14\%  of 
carriers in HTS [1] and 4\% in Pb [29]  at T = 0 and H = 0, and  may have a maximum value 1/2 
at $<N_s>~ = 1/2$.   For a given  $T, ~B(T, H)/H_{c2}$  varies from 0 at  $H = H_{c1}$  to 1 at  $H_{c2}$. Eq. (19B) 
becomes $-W_2 h_2$ at  $H = H_{c1}$.   For a  given $H,~ B(T,  H) / H_{c2}$ varies from  $B(0, H)/H_{c2}$  to $B(T_c, 
H)/H_{c2} = H/H_{c2}$.  For YBCO of [15],  at $T = 0, ~B(0, H)/H_{c2} <  H/H_{c2} = 6/100, ~2S = 2\times 0.14 \times 0.86,$ 
then Eq. (19B) is negative for $0.25 <  Ph_2/h_1$.  Thus, both cases of T and  H sweepings, can have 
the sign changes of the  Hall field three times,  since S is not  a monotonic function of  T and H.  
Physically, the Lorentz force [6]  may be responsible for VF  motion and the Magnus  force [7] for 
VAF, since  VF [VAF]  has  the magnetic  flux greater   [less] than the  effective  flux via  the 
circulating current.  Then, $h_1 = ~<B>$ and $h_2 = 1$, Eq. (19B) varies as $<B> - ~2PS/<B>$ for $\tau_2/\tau_1 > 
2PS/<B>~ =  n_2/n_1$,  compatible with   data [12a].  From   Eqs. (18)  and (21),   we may get   the 
resistivity/the normal state resistivity $\rho_n$ as
\setcounter{equation}{21}
\begin{eqnarray}
\rho_{xx}/\rho_n = ~< B > \tau /\tau_1 + 2PS\tau /\tau_2,
\end{eqnarray}
where $\tau$ is the carrier relaxation time in the normal state.  From Eqs. (19B) and (22), we may get 
the scaling behavior of the Hall resistivity as 
\begin{eqnarray}
\rho_{yx} \propto \rho^2_{xx}[ <B> h_1 - 2PS h_2 ]/ [<B> \tau / \tau_1 + 2PS \tau /\tau_2]^2.
\end{eqnarray}
Without VAF terms, Eq. (23) becomes $\rho_{yx} \propto  \rho^2_{xx}$ [11] for $h_1 = ~<B>$ and $\tau = \tau_1$ [6].

In summary, the  notion of  MS [3]  is essential to  have three  magnetic quantum  states of 
vortices.  VZF is a  new vortex quantum  state and may act  as a domain  wall between VF  and 
VAF regions, accounting for the observation  of the flux and antiflux regions  in Nb films [26].  A 
direct detection of VZF may be hard as that of neutral particles.  VF and  VAF numbers are found 
to play major roles for the sign changes of the Hall field, even though the present analysis is based 
on the basic Eq. (20) of the vortex core models. In  general, Eqs. (18) and (19) may be used.  The 
notion of superconducting ringlets may account for the paramagnetic Meissner effect [22].  The fact 
that three vortices can account for  data [26] and the Hall anomaly,  indicates the present theory is 
sound. 
 
I thank John  R. Clem  for Ref.  [23] and  Ju H.  Kim for  his interest  in this  problem by 
obtaining the Hall angles [30] quantitatively  in good agreements with data  of HTS and LTS, via 
modeling PS functions.
\newpage
{\bf References}

[1] S. B. Nam, Phys. Lett. A 193 (1994) 111; (E) A 197 (1995) 458.

[2] J. Bardeen, L. N. Cooper, and J. R. Schrieffer, Phys. Rev. 108 (1957) 1175.

[3] S. B. Nam, Phys. Lett. A 198 (1995) 447.

[4] M. R. Beasly et al, Phys. Rev. Lett. 42 (1979) 1165.

\hspace*{0.6cm}S . Doniach and B. A. Hubermanm, Phys. Rev. Lett. 42 (1979) 1169.

\hspace*{0.6cm}B. I. Ivlevvet al, Phys. Rev. B 52 (1995) 13532.

[5] J. M. Kosterlitz and D. J. Thouless, J. Phys. C 6 (1973) 1181.

\hspace*{0.6cm}V. L. Berezinskii, JETP 32 (1971) 493; 34 (1972) 610.

[6] J. Bardeen and M. J. Stephen, Phys. Rev. 140 (1965) A1197.

[7] P. Nozieres and W. F. Vinen, Philos. Mag. 14 (1966) 667.

[8] P. Ao and D. J. Thouless, Phys. Rev. Lett. 70 (1993) 2158.

[9] Y. Iye et al. Physica C 167 (1990) 278.

[10] Z. Hao, C-R. Hu, and C. S. Ting, Phys. Rev. B 51 (1995) 9387.

[11] V. M. Vinokur et al, Phys. Rev. Lett. 71 (1993) 1242. 

[12] J. Hagen et al, Phys. Rev. B 41 (1990) 11630 ; B 47 (1993) 1064.

\hspace*{0.8cm}(a) J. M. Harris et al, Phys. Rev. B 51 (1995) 12053.

\hspace*{0.8cm}D. M Ginsberg and J. T. Manson, Phys. Rev. B 51 (1995) 515.

\hspace*{0.8cm}Y. Matsuda et al , Phys. Rev. B 52 (1995) R15749.

[13] M. N. Kunchur et al, Phys. Rev. Lett. 72 (1994) 2259.

\hspace*{0.8cm}A. V. Samoilov et al, Phys. Rev. Lett. 74 (1994) 2351.

\hspace*{0.8cm}The recent observation of the pining dependence of the Hall field in YBCO with 

\hspace*{0.8cm}columnar defects by W. N. Kang et al, Phys. Rev. Lett. 76 (1996) 2993 may be 

\hspace*{0.8cm}understood as the scattering effects ($\tau_i$) in the 
present theory via Eqs (19A), (22), 

\hspace*{0.8cm}and (23).

[14] J. M. Graybeal et al, Phys. Rev. B 49 (1994) 12923.

\hspace*{0.8cm}A. W. Smith et al, Phys. Rev. B 49 (1994) 12927.

[15] B. Parks et al, Phys. Rev. Lett. 74 (1995) 3265.

[16] Z. D. Wang et al, Phys. Rev. Lett. 67 (1991) 3618; 72 (1994) 3875.

[17] R. A. Ferrell, Phys. Rev. Lett. 68 (1992) 2524.

[18] D. I. Khomskii and A. Freimuth, Phys. Rev. Lett. 75 (1995) 1384.

[19] A. T. Dorsey, Phys. Rev. B 46 (1992) 8376.

[20] A. van Otterlo et al, Phys. Rev. Lett. 75 (1995) 3736.

[21] A. A. Abrikosov, JETP 5 (1957) 1174.

[22] S. B. Nam, to be published.

[23] J. R. Clem. J. Low Temp. Phys. 18 (1975) 427.

[24] L. Neuman and L. Tewordt, Z. Physik 189 (1966) 55.

[25] W. A. Little and R. D. Parks, Phys. Rev. Lett. 9 (1962) 9.

[26] C. A. Duran et al, Phys. Rev. B 52 (1995) 75.

[27] J. R. Kirtley et al, Phys. Rev. B 51 (1995) 12057.

[28] B. D. Josephson, Phys. Lett. 16 (1965) 242.

[29] S. B. Nam, J. Korean Phys. Soc. 28 (1995) 102.

[30] J. H. Kim and S. B. Nam, to be published.
\end{document}